\documentclass[fleqn,usenatbib]{mnras}
\usepackage[T1]{fontenc}
\usepackage{ae,aecompl}
\usepackage{graphicx}   % Including figure files
\usepackage{amsmath}    % Advanced maths commands
\usepackage{amssymb}    % Extra maths symbols

\usepackage[dvipsnames]{xcolor}

\title[Impulsive SFE]{An impulsive geomagnetic effect from an early-impulsive flare}

\author[H.S. Hudson et al.]{
\begin{tabular}{lll}
Hugh S. Hudson,$^{1,2}$\thanks{hugh.hudson@glasgow.ac.uk}, Edward W. Cliver$^{3}$, Lyndsay Fletcher$^{1,4}$, Declan A. Diver$^{1}$, \\ Peter T. Gallagher$^{5}$, Ying Li$^{6}$, Christopher M. J. Osborne$^{1}$, Craig Stark$^{1}$,Yang Su$^{6}$
\end{tabular}
\bigskip
\\
$^{1}$SUPA School of Physics and Astronomy, University of Glasgow, Glasgow G12 8QQ UK\\
$^{2}$Space Sciences Laboratory, University of California, Berkeley, CA 94720 USA\\
$^{3}$National Solar Observatory, 3665 Discovery Drive, Boulder CO 80303 USA\\
$^{4}$Rosseland Centre for Solar Physics, University of Oslo, PO Box 1029 Blindern, NO-0315 Oslo, Norway\\
$^{5}$Astronomy \& Astrophysics Section, DIAS Dunsink Observatory, Dublin Institute for Advanced Studies, Dublin, D15 XR2R, Ireland\\
$^{6}$Key Laboratory of Dark Matter and Space Astronomy, Purple Mountain Observatory, Chinese Academy of Sciences,\\ Nanjing 210034, China
}

\date{Accepted XXX. Received YYY; in original form ZZZ}
\pubyear{WWW}

\begin{document}

\label{firstpage}
\pagerange{\pageref{firstpage}--\pageref{lastpage}}
\maketitle

\maketitle

\begin{abstract}
The geomagnetic ``solar flare effect'' (SFE) results from excess ionization in the Earth's ionosphere, famously first detected at the time of the Carrington flare in 1859.
This indirect detection of a flare constituted one of the first cases of ``multimessenger astronomy,'' whereby solar ionizing radiation stimulates ionospheric currents.
Well-observed SFEs have few-minute time scales and perturbations of $>10$~nT, with the greatest events reaching above 100~nT.
In previously reported cases the SFE time profiles tend to resemble those of solar soft X-ray emission, which ionizes the D-region; there is also a less-well-studied contribution from Lyman $\alpha$.
We report here a specific case, from flare SOL2024-03-10~(M7.4), 
in which an impulsive SFE deviated from this pattern.
This flare contained an ``early impulsive'' component of exceptionally hard radiation, extending up to $\gamma$-ray energies above 1~MeV, distinctly before the bulk of the flare soft X-ray emission.
We can characterize the spectral distribution of this early-impulsive  component in detail, thanks to the modern extensive wavelength coverage.
A more typical gradual SFE occurred during the flare's main phase.
We suggest that events of this type warrant exploration of the solar physics in the ``impulse response'' limit of very short time scales.
\end{abstract} 

\begin{keywords}
Sun: EUV -- Sun: X-rays -- Sun: corona
\end{keywords}

\section{Introduction}\label{sec:intro}
The first reported optical observation of a flare \citep{1859MNRAs..20...13C} coincided with the first recorded ``solar flare effect'' (SFE) \citep{1861RSPT..151..423S}.
It appeared in all three components of the geomagnetic field in close time coincidence with the flare's optical emission, andhad a magnitude of about 120~nT (of order 1\% of the background geomagnetic field) mainly in the horizontal component of the field.
Since this time hundreds of SFEs have been detected, as reviewed recently by \cite{2020JSWSC..10...27C}.
The SFE for the Carrington flare remains topically important  in view of the dangers posed by extreme space-weather events \citep[e.g.,][]{2022LRSP...19....2C}.

Any terrestrial magnetometer on the sunlit side of the Earth will detect an SFE of sufficient magnitude to outweigh the complicated natural variability of the geomagnetic field.
The effects on the vector field at a given station depend upon its location relative to the sub-solar point, as well as other factors \citep{2020JSWSC..10...27C}.
At middle latitudes in the northern hemisphere, as with the Carrington flare, the appearance of a typical SFE is a smooth depression of the horizontal component generally following the time evolution of the solar soft X-ray flux.
\cite{2020JSWSC..10...27C} describes the complexity of the SFE detection and morphology in greater detail.

The recent event SOL2024-03-10~(GOES class\footnote{This adheres to the current NOAA calibration \citep[e.g.][]{2024SoPh..299...39H}.} M7.4) displays the usual pattern of an SFE, in that it has a slowly-developing ionospheric component roughly coinciding with the soft X-rays. 
It is strikingly different, though, in that it also shows an early sharp component in response to an ``early impulsive'' hard X-ray burst. 
In early impulsive flares,  the $>$25~keV hard X-ray flux increase is delayed by less than 30~s after the soft X-ray onset \citep{2007ApJ...670..862S}.
Such flares can depart from the observed close relationship found in the common Neupert effect \citep{1968ApJ...153L..59N,1991BAAS...23R1064H,1993SoPh..146..177D}.
The Neupert effect underpins standard flare models by linking the energy in accelerated particles to that required for injecting hot plasma into coronal flare loops.

\begin{figure}
\centering
    \includegraphics[width=0.49\textwidth]{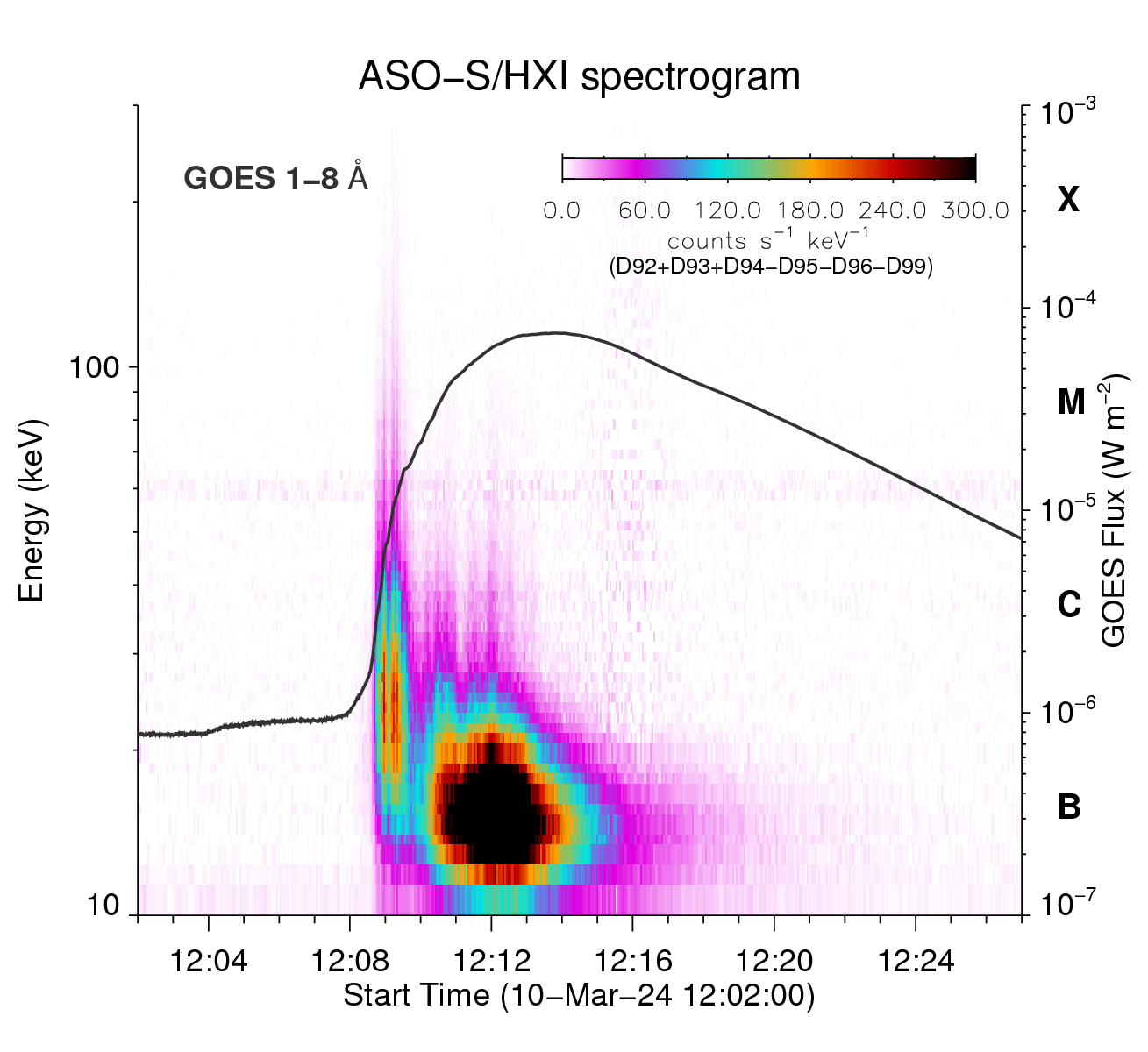}
 \caption{Background-subtracted ASO-S X-ray spectrogram of SOL2024-03-10, showing the early-impulsive phase at 12:08-12:10~UT. 
The legend indicates the HXI detectors used for the signal and for the background corrections.
We interpret the 10-20~keV X-ray response as the thermal soft X-rays of the gradual phase, and the spikes at 30 keV over 12:10--12:13~UT as the normal impulsive phase.
The black line shows the GOES 1-8~\AA\ flux.
}
\label{fig:SOL2024-03-10_ASO-S}
\end{figure}

The simplest description of flare time development consists of three phases: onset, impulsive, and gradual, as recognized in soft X-ray emissions.
The onset phase \citep[e.g.,][]{2021MNRAS.501.1273H} typically is faint and not relevant to SFE formation; the impulsive phase \citep{1970ApJ...162.1003K} includes non-thermal hard X-ray and microwave signatures, while the gradual phase consists of hot plasma injected into the corona, and the well-known flare loops \citep{1968ApJ...153L..59N}.
The ``early impulsive'' character of the event described here falls outside this standard pattern because it precedes the normal impulsive development, which we identify with the Neupert effect.
This distinctive property is immediately evident in the hard X-ray spectrogram from the ASO-S/HXI instrument \citep{2019RAA....19..160Z,2019RAA....19..163S} on board ASO-S \citep{2023SoPh..298...68G}, as seen in Figure~\ref{fig:SOL2024-03-10_ASO-S}.

In Section~\ref{sec:obs} we survey some of the databases containing information about the early-impulsive epoch of SOL2024-03-10, describing a broad spectral energy distribution (SED) covering some six decades from the UV to the MeV $\gamma$-ray range. 
We note that at the high-energy end of the observed SED, the photons penetrate the Earth's atmosphere to heights well below the ionospheric D-region.
This layer absorbs most of the solar soft X-ray and Lyman-$\alpha$ fluxes \citep[e.g.,][]{1974ASSL...46.....M}.
The $\gamma-ray$ photons cause multiple ionizations via secondary processes. 
The ``specific ionization'' approximation \citep[e.g.,][]{1950nuph.conf.....F} implies roughly 30~eV per ion pair; a single MeV photon will thereby produce more than 10$^4$ secondary ions in the mesosphere  and down into the stratosphere at altitudes of 30-40~km.
\cite{2002JPhD...35.1311M} describes the physics of a $\gamma$-ray component, which he terms a ``crotchet\footnote{Often spelled ``crochet'' (the French word for ``hook''), an early descriptive term for an SFE.} impulse component'' or CIC. 

\section{Observations}\label{sec:obs}

\subsection{Overview of magnetic records}

\begin{figure}
\centering
    \includegraphics[width=0.49\textwidth]{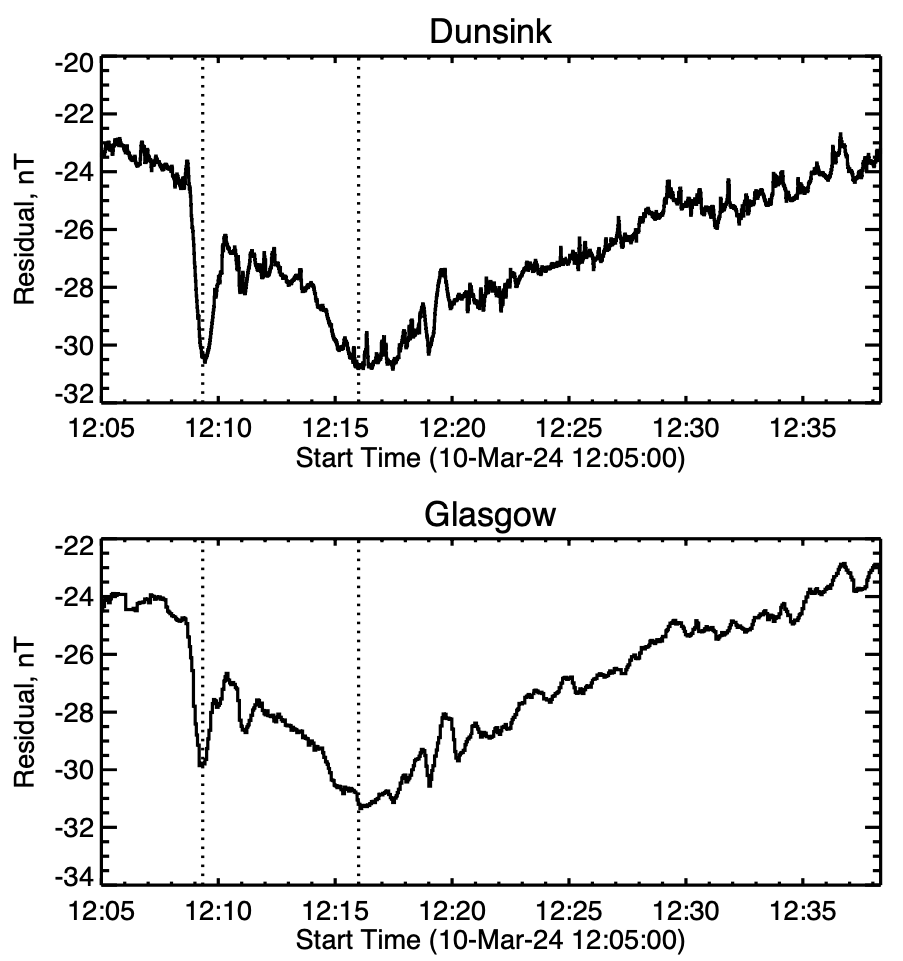}
 \caption{Comparison of magnetometer records from Dunsink (1~sec sampling) and Glasgow (3~sec sampling) for SOL2024-03-10, showing residuals against daily median values for the horizontal component $B_{hor}$. 
 The dotted lines show approximate maximum times for the fast SFE component (12:09:20~UT) and the normal SFE (12:16~UT) respectively.
}
\label{fig:simple_crotchet}
\end{figure}

The subject flare SOL2024-03-10T12~(M7.4) occurred at S13W38 in NOAA AR~13599 and its SFE immediately marked it as unusual, as confirmed directly by the hard X-ray spectrogram made available by the ASO-S database in near real time (Figure~1).
Many spacecraft and observatories detected various aspects of this event.
We use magnetometer data at 1-sec cadence from Dunsink Observatory, nearly on the meridian of the subsolar point at about 20$^\circ$ to the S, and check with 3-s data from Glasgow {\b University} Observatory\footnote{\url{www.magie.ie}}, with the comparison shown in Figure~\ref{fig:simple_crotchet}.
Our two geomagnetic datasets come from Dunsink (N 53$^\circ$ 23$'$ 13.40$''$, W 6$^\circ$ 20$'$ 15.17$''$) and Glasgow (N 55$^\circ$ 51$'$ 26$''$, W 4$^\circ$ 15$'$ 7$''$).

Generally this event had a stronger geomagnetic effect ($\Delta B_{hor} \approx$ 7~nT)  than other flares of this GOES class.
A  distinguishing feature of this SFE is its fast component at about 12:09~UT, marked with the left-hand dotted line in Figure~\ref{fig:simple_crotchet}.
The data agree reasonably well and have correlated irregular background variations at the nT level.
The Glasgow 3-s data are smoother than the Dunsink 1-s data, as would be expected from the longer integrations.
Similar two-step SFE time profiles were recorded at Eskdalemuir and Hartland in the UK\footnote{\url{https://imag-data.bgs.ac.uk/GIN_V1/GINForms2}.}.

\subsection{Hodogram representation}

\begin{figure*}
\centering
    \includegraphics[width=\textwidth]{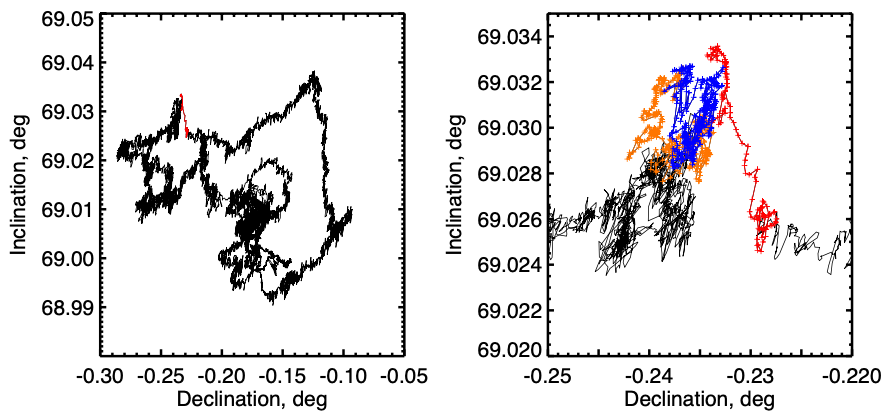}
 \caption{Hodogram plots of geomagnetic inclination (zenith = 90$^\circ$) against azimuthal angle, measured from N to E for the Dunsink data.
These are the geomagnetic inclination and declination, respectively.
The left panel shows the full day, during which the field executes a complicated counter-clockwise motion  typical of the diurnal effect at this geographical point.
The right panel focuses on the flare interval so that one can identify the excess due to the 
SFE.
The colored time ranges separate 12:08 (red), 12:10 (orange), 12:16 (blue), up to 12:22~UT.
Note the different deflection patterns for the early-impulsive component (red) and the main SFE.
}
\label{fig:hodo_2up}
\end{figure*}

During the SFE we can follow the disturbance vector via a hodogram, as shown in Figure~\ref{fig:hodo_2up} here.
This shows the angular motions of the total geomagnetic vector.
The two panels show the diurnal circulation of the Dunsink geomagnetic vector and a blown-up region showing the time interval of the impulsive SFE component.
The diurnal motion of the direction of ${\bf B}$ shows a counterclockwise pattern.
The details of the SFE motions have interesting differences: the impulsive component shows a tilt to the N at onset, followed by a return on approximately the same path.
Then, as shown by the colors in the right panel of Figure~\ref{fig:hodo_2up}, the field again exhibits a rotation, but roughly in the opposite sense (clockwise) from that of the diurnal effect.

The hodogram suggests (from the Biot-Savart law) that the initial impulse in ${\bf B}$ results from a westward horizontal current if driven from the lower latitudes nearer the subsolar point, which for this flare lies approximately on the same meridian as the magnetic observatories.

\subsection{Neupert Effect}

\begin{figure*}
\centering
    \includegraphics[width=\textwidth]{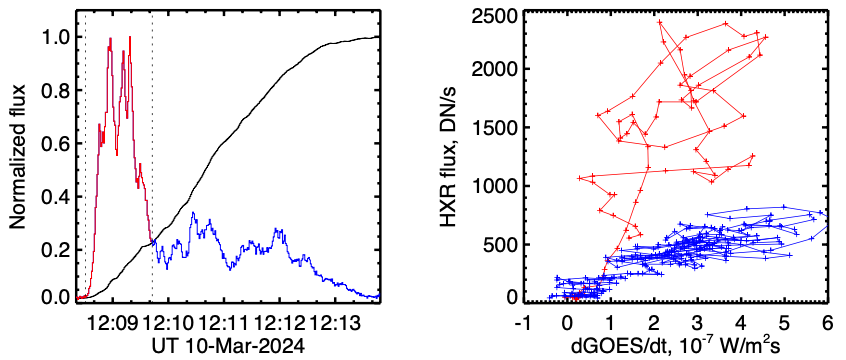}
 \caption{Left panel, time series of Fermi/GBM hard X-ray (31~keV) and GOES soft X-ray (1-8~\AA) fluxes up to the time of GOES maximum.
The right panel shows the correlation between the hard X-ray flux and the time derivative of the GOES flux at 1-sec sampling, with the early-impulsive phase points in red and the later points in blue.
Both sets of points show strong correlations.
}
\label{fig:early_impulsive}
\end{figure*}

The event importantly displays a clear distinction between the early-impulsive component and the normal Neupert-effect component, as shown in the time-series comparison in Figure~\ref{fig:early_impulsive} (left panel).
This simplifies the relationship easily seen in the ASO-S hard X-ray spectrogram (Figure~\ref{fig:SOL2024-03-10_ASO-S}), but now relating the timing to the GOES/XRS soft channel (1-8~\AA).

The right panel of Figure~\ref{fig:early_impulsive} directly compares the time derivative of soft X-rays with the hard X-ray flux (\textit{Fermi}/GBM NaI detector n0, energy channel 22).
This derivative comparison is complementary to Neupert's original integral version, which related the time integral of impulsive microwave emission to the soft X-ray time history.
Both sets of points indeed correlate well, with Pearson's $\rho = [0.77, 0.85]$ respectively for the early-impulsive and impulsive phases.
The clear distinction of the two sets of points implies that the early-impulsive particle acceleration drives a significantly weaker rate of mass transfer (evaporation) into the corona.

\subsection{Spectral Energy Distribution}

\begin{figure}
\centering
    \includegraphics[width=0.49\textwidth]{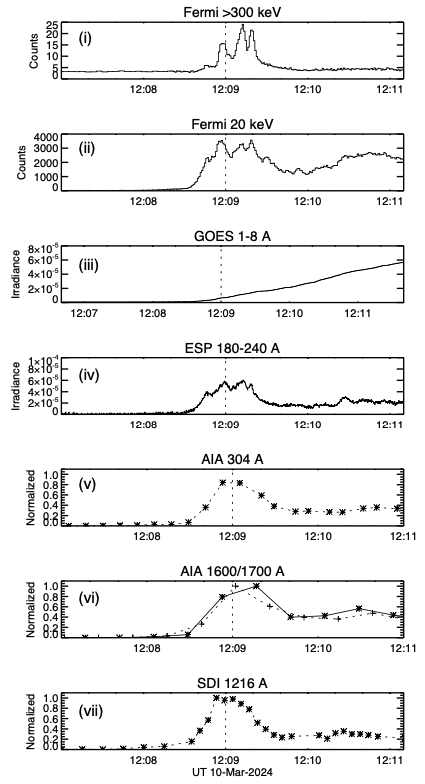}
 \caption{Data timeseries across a wide spectral range. The top two panels show \textit{Fermi}/GBM in counts; the GOES and ESP data are in irradiance units (W/m$^2$), while the lower three panels show excess fluxes normalized to the event peak.
The vertical dotted line shows a reference time for the early-impulsive emissions.
}
\label{fig:broad_ts}
\end{figure}

As seen, the isolated fast SFE corresponds to an unusual ``early impulsive'' flare emission.
In this section we describe the spectral energy distribution via time series across a broad spectral range.
Figure~\ref{fig:broad_ts} shows representative data, ordered
top to bottom in decreasing photon energy:
\begin{enumerate}
\item \textit{Fermi}/GBM counting rates above 300~keV: the $\gamma$-ray continuum.
This extends up to at least 1.5~MeV.
\item \textit{Fermi}/GBM counting rates at 20~keV, the more typical hard X-ray continuum defining the main impulsive phase.
\item GOES XRS-B (1-8 \AA, or roughly 2 keV): the hot thermal plasma in flare loops.
\item SDO/EVE/ESP (18~nm, nominally 0.15 keV): the EUV line forest from flare loops and footpoint regions.
\item SDO/AIA 304~\AA\ (nominally 40~eV): largely chromospheric.
\item SDO/AIA 1600 and 1700~\AA~(nominally 7.5 eV): The corona/chromosphere interface region \citep{2019ApJ...870..114S}.
\item ASO-S/SDI Lyman-$\alpha$~(around 10~eV): largely chromospheric.
\end{enumerate}

Note that coronal thermal emissions dominate the soft X-ray range (iii).
The line emissions in items (iv--vii) all show the impulsive character of the hard X-rays, however, with time scales for flux variation $\tau = S/{\dot S}$ as short as 5~s.
The Lyman-$\alpha$ line is known to have both impulsive and gradual signatures in other flare events \citep{2020SpWea..1802331M}.
Notably the GOES/SXR data do not show rapid fluctuations directly because of the slowly-cooling coronal hot plasma of the flare loop system \citep[e.g.,][]{1982ApJ...263..409A}.

The multiwavelength timeseries data for the early-impulsive phase (Figure~\ref{fig:broad_ts}) reveal two important properties.
First, all wavebands excepting the soft X-rays from GOES have quite similar time profiles despite their different emission mechanisms.
Second, the Figure does not include a plot of visible continuum (the white-light flare).
We have examined the HMI pseudocontinuum, which normally shows white-light flares well, and see only faint and diffuse emissions with little energy and no hint of the fine spatial structures seen in the EUV imaging (Section~\ref{sec:imaging}).
The left panel of Figure~\ref{fig:early_impulsive} shows the X-ray time development for the event, comparing normalized signals for one Fermi/GBM NaI detector at about 31~keV, and the standard GOES-16/XRS 1-8~\AA\ data.
The hard X-ray timing matches that of the fast SFE quite well, but the maximum (negative) excursion of the slow SFE is delayed by about 3 minutes from the 1-8~\AA\, suggesting that the main driver of this later component is at longer wavelengths.

\subsection{Time Variability}

\begin{figure}
\centering
    \includegraphics[width=0.49\textwidth]{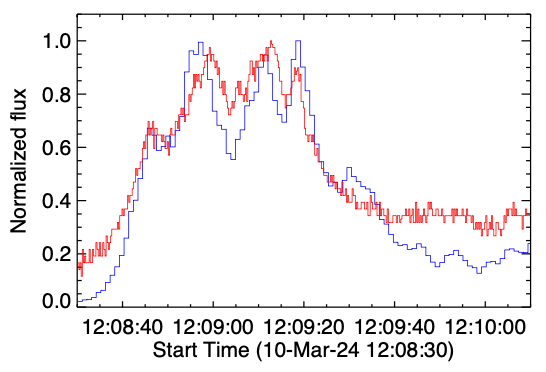}
 \caption{Hard X-rays (31~keV at 1~s binning), blue, compared with EUV 18-24~nm at 1/4~s binning, red.
}
\label{fig:timelag}
\end{figure}

At the longest wavelengths the available data do not have sufficient time resolution to follow the hard X-ray variability, but the ESP broadband EUV time-series data (item \textit{iv} in Figure~\ref{fig:broad_ts}) capture this with 1/4~s sampling \citep{2012SoPh..275..179D}.
Accordingly, because of its relevance to understanding causality in the solar sources of the radiation, we compared \textit{Fermi} 31-keV with the ESP 18-24~nm during the early impulsive phase in Figure~\ref{fig:timelag}.
This gives ambiguous conclusions: of the five major peaks in hard X-rays, the EUV appears to match to within about a second for the three peaks at about 12:08:45, 12:09:10, and 12:09:18~UT;  but the ESP peak at 12:09:00~UT clearly lags the hard X-rays by 2-3~s.
The final hard X-ray peaks after about 12:09:25 have no detectable counterparts in the EUV at all.

The early-impulsive SFE has the approximate duration of the multispectral emissions, but may not have sufficient signal-to-noise ratio to show the fine structure seen in the hard X-ray and EUV bands.

\subsection{Flare imaging}\label{sec:imaging}

\begin{figure*}
\centering
    \includegraphics[width=\textwidth]{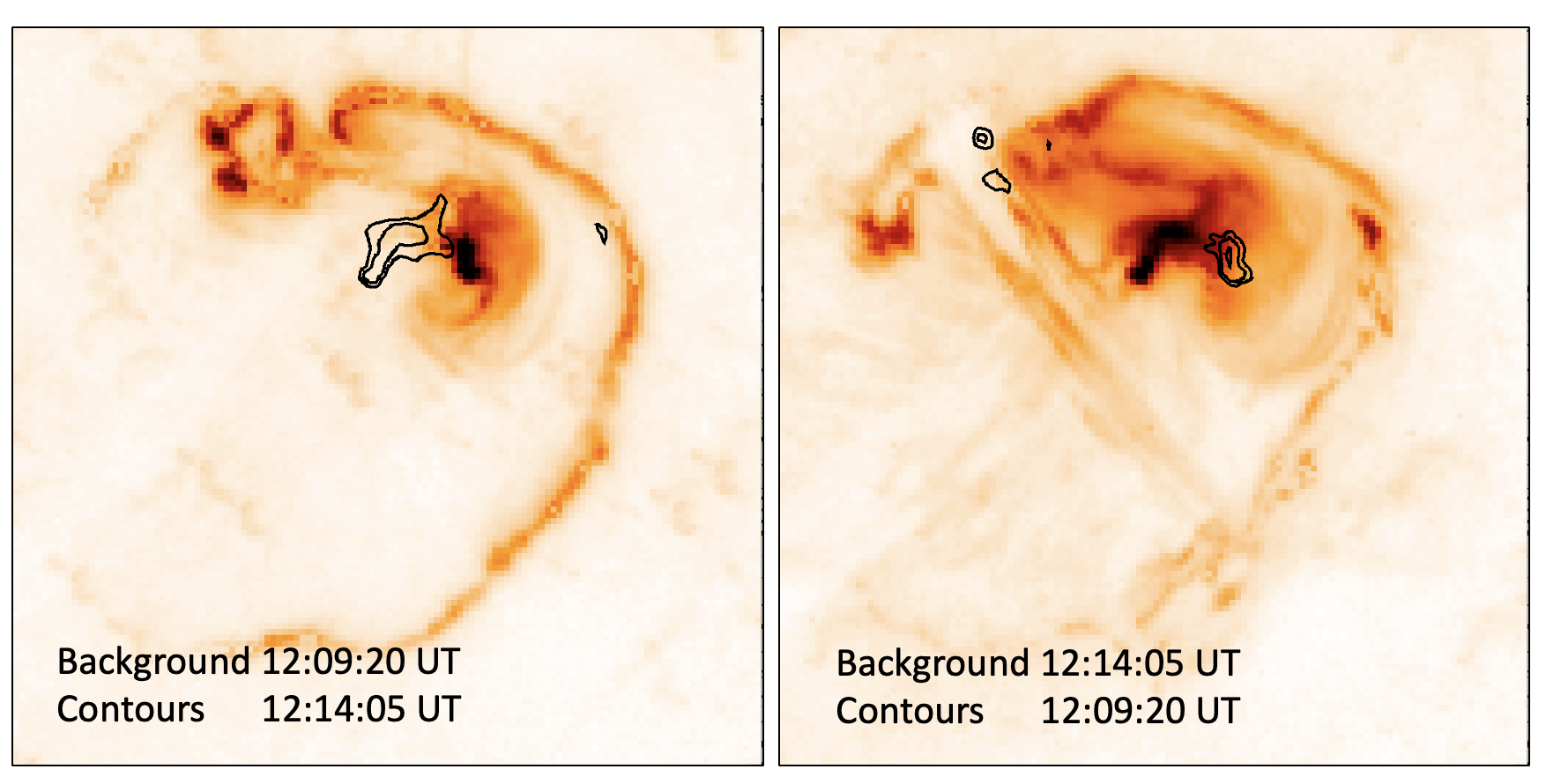}
 \caption{AIA 304~\AA\ images ($64'' \times 64''$) of SOL2024-03-10 at 12:09:20~UT, just at the end of the early-impulsive peak (color), and a main-phase image at 12:14:05~UT.
The images are reversed-color and independently scaled by the square root of the brightness.
Left, the early-impulsive image, showing compact footpoints, with contours from the main-phase image;
right, the reverse view. 
The large-scale arc, which we suggest corresponds to the dome of the spine-and-fan geometry, does not move greatly as the flare develops, and the main-phase emission concentrates between the early-impulsive footpoints (see text for commentary).
}
\label{fig:SOL2024-03-10_image}
\end{figure*}

The detection of the early-impulsive process at EUV wavelengths allows us to locate the footpoint regions of the flaring structure.
We sketch its relevant properties with the overlay in Figure~\ref{fig:SOL2024-03-10_image} here.
This shows a cutout region of the AIA 304~\AA\ image at 12:09:20~UT (peak pixel brightness $\approx 2.6 \times 10^4$~DN/sec), overlaid with an image at 12:14:05~UT (peak pixel brightness $\approx 5 \times 10^3$~DN/sec).
The two panels show the two images, each with the other as contours for ready comparison of the geometry.
We list some of the key features:
\begin{itemize}
\item The early-impulsive image consists of compact brightenings that are basically unresolved, and thus contain sub-arcsecond structure.
\item The early-impulsive brightenings occur in two regions, in a widespread double-footpoint pattern.
\item These bright points are unresolved in time by the AIA image cadences.
\item The gradual-phase image contours lie interior to the early-impulsive footpoints and develop a two-ribbon structure.
\item The prominent outer circular-ribbon structure evolves only slowly, not moving significantly.
\end{itemize}

This complicated development contains many clues to the flaring structure.
The configuration suggests spine-and-fan connectivity for the magnetic field, with the outer structure identifiable with the intersection of the dome and the photosphere \citep[][]{2001ApJ...554..451F,2009ApJ...700..559M,2012A&A...547A..52R}.
In this interpretation, the images suggest that the main flare development lies inside (below) the preflare dome, while the western  early-impulsive footpoint could show the location of the interior spine.

\section{Discussion}

The early-impulsive flare SOL2024-03-10 gives us a chance to characterize an unusual perturbation of the Earth's upper atmosphere via its observed spectral energy distribution.
This suggests contributions across a broad altitude range in the upper stratosphere and mesosphere, and this ``test pulse'' distinctly differs from the usual SFE behavior both in the time domain and in the spectrum.
It should therefore make an interesting exercise for model development of the current system \citep{2002JPhD...35.1311M}.
We note that the hodogram analysis reveals a different pattern than for the subsequent main phase of the SFE, possibly implying some feedback in the development of the enhanced currents.

Similarly, isolating the early-impulsive solar emission, with its relatively hard spectrum, should interest modelers of flare development
The relative weakness of evaporation for a harder primary electron spectrum results from the greater depth of penetration of higher-energy electrons, which can be modeled by 1D radiation hydrodynamics \citep[e.g.][]{2015ApJ...808..177R}.
The major puzzle presented by this event is its lack of white-light continuum, which conventionally should result from excitation in the deep solar atmosphere \citep[e.g.,][]{2012ApJ...753L..26M}.

The impulsive SFE appears most strongly in the geomagnetic X component (N pointing).
According to the Biot-Savart law, this corresponds to a westward horizontal current,
consistent with a flare subsolar point on the meridian \citep[e.g.][]{2020JSWSC..10...27C}.
In the subsequent development of the normal SFE component here, the hodogram shows a similar counter-clockwise circulation as seen in the diurnal motion at Dunsink and Glasgow on the day of this event.
It is beyond the scope of this paper to study this SFE in detail, but many geomagnetic observatories will have records of it at high time resolution.

\section{Conclusions}

We have reported on the geomagnetic effects of a recent flare exhibiting a rapid ionospheric response to an ``early impulsive'' flare emission.
The spectral energy distribution for the solar emissions extend to the $\gamma$-ray energy range.
There are other such two-step SFE events, presumably with to early $\gamma$-ray emission, such as SOL1984-04-24 \citep{2020ApJ...903...41C}, and so the mechanism isolated in our case is likely not unique.
Future modeling of SFE development should  benefit from this new kind of test pulse, with its reduced confusion from the usual soft X-ray component.

``Early impulsive'' behavior is not rare \citep{2007ApJ...670..862S}, but we have shown that such a burst may result in a deviation from the simplest form of the Neupert effect \citep[e.g.][]{2005ApJ...621..482V}.
In this case we find that the early-impulsive emissions appear in compact footpoint regions, presumably connected by large-scale fields in a spine-and-fan geometry.

\section{\bf Acknowledgements:} 
We thank the anonymous reviewer for helpful comments.
The recognition of the event we have studied was facilitated by the availability of ASO-S near-real-time data via \url{http://sprg.ssl.berkeley.edu/~tohban/browser/}, for which we thank Albert Shih. 
ASO-S  is supported by the National Key R\&D Program of China 2022YFF0503002, the Strategic Priority Research Program of the Chinese Academy of Sciences, Grant No. XDB0560000, the NSFC 12333010
and NSFC 11820101002.  
L. F. acknowledges support from grant ST/X000990/1 made by the UK's Science and Technical Facilities Council.
H. S. H. thanks the University of Glasgow for hospitalilty, and Graham Woan for great assistance with the Glasgow magnetometer.
For the purpose of open access, the authors have applied a Creative Commons Attribution (CC BY) licence to any Author Accepted Manuscript version arising from this submission.

\section{\bf Data Availability Statement:} 

All data used in this article are publicly available.
We thank the Magnetometer Network of Ireland (\url{https://www.magie.ie}) for the geomagnetic record used here.

\bibliographystyle{mnras}
\bibliography{sfe}

\end{document}